\documentclass[%
 aip,
 amsmath,amssymb,
 reprint,%
]{revtex4-1}

\usepackage{graphicx}
\usepackage{dcolumn}
\usepackage{bm}
\usepackage[mathlines]{lineno}
\usepackage[utf8]{inputenc}
\usepackage[T1]{fontenc}
\usepackage{mathptmx}
\usepackage{etoolbox}

\makeatletter
\def\@email#1#2{%
 \endgroup
 \patchcmd{\titleblock@produce}
  {\frontmatter@RRAPformat}
  {\frontmatter@RRAPformat{\produce@RRAP{*#1\href{mailto:#2}{#2}}}\frontmatter@RRAPformat}
  {}{}
}%
\makeatother
\begin{document}

\preprint{AIP/123-QED}

\title{Electronic structure and defect properties of Bi-doped GaN: origins of photoluminescence and optical absorption}

\author{Yujie Liu}
\email{yujl@umich.edu}
\affiliation{Department of Materials Science and Engineering, University of Michigan, Ann Arbor, Michigan 48109, USA}

\author{Ishtiaque Ahmed Navid}%
\affiliation{Department of Electrical Engineering and Computer Science, University of Michigan, Ann Arbor, 1301 Beal Avenue, Ann Arbor, Michigan 48109, USA}

\author{Zetian Mi}
\affiliation{Department of Electrical Engineering and Computer Science, University of Michigan, Ann Arbor, 1301 Beal Avenue, Ann Arbor, Michigan 48109, USA}

\author{Emmanouil Kioupakis}
\affiliation{Department of Materials Science and Engineering, University of Michigan, Ann Arbor, Michigan 48109, USA}

\date{\today}

\begin{abstract}
Extreme lattice-mismatched III–V nitrides, such as Bi-incorporated GaN, have been realized experimentally thanks to recent advances in epitaxial growth and characterization techniques. However, theoretical insights into defect-related optical absorption and emission phenomena in these materials remain scarce. Here, we apply hybrid density functional theory to systematically explore the role of substitutional bismuth atoms on both cationic ($\rm Bi_{Ga}$) and anionic ($\rm Bi_N$) sites in Bi-incorporated GaN, as well as their complexes with native vacancies. Our calculations reveal that the charge-compensated $(\rm Bi_N + V_{Ga})^{3-}$ and $(\rm Bi_N + V_{Ga})^{3+}$ defect complexes stabilize anionic bismuth incorporation, accounting for the experimentally observed absorption peaks at ~1.11 eV and ~3.17 eV. We further uncover the origins of the reported band-edge emissions near ~2.0 eV and ~2.5 eV by examining various charge states of $\rm Bi_{Ga}$ and $\rm Bi_N$ centers. Our findings elucidate the defect-level physics of Bi-doped GaN and provide practical guidelines for controlling the incorporation of Bi into GaN.   
\end{abstract}

\maketitle

The incorporation of bismuth (Bi) into gallium nitride (GaN) has drawn significant attention both for thin films\cite{novikov2003bismuth, foxon2002bismuth, zhang2002influence} and for nanostructures \cite{navid2024structural} due to its intrinsic highly mismatched atomic size and ionicity. These alloys exhibit intriguing cathodoluminescence (CL), Raman scattering,\cite{ibanez2005optical} photoluminescence (PL), and optical absorption characteristics arising from the interplay between the localized Bi states and the host GaN lattice.\cite{kudrawiec2020bandgap} These unique properties, such as a redshift in emission and enhanced absorption in the near-infrared (NIR) region,\cite{novikov2003bismuth} have made Bi-doped GaN highly promising for applications in optoelectronic devices, including light-emitting diodes (LEDs), lasers, photodetectors, etc. However, the detailed interactions between the incorporated Bi and the host GaN material remain unclear. An understanding of the atomistic origins of these optical phenomena is essential for optimizing these materials and realizing their potential in devices.

Overall, three as-yet unexplained experimental results about the band gap of Bi-doped GaN have been previously reported in the literature. Experiments by Levander \textit{et al.}\cite{levander2010gan1} and Novikov \textit{et al.}\cite{novikov2012molecular} reported that Bi alloying in the GaN lattice at the dilute limit ($< 5\%$ Bi concentration) results in band-edge PL peak or optical absorption onset at around 3.2 eV, i.e., near the band gap of GaN. In both cases, the Ga$\rm N_{1-x}Bi_x$ films were grown on sapphire substrates by plasma-assisted molecular beam epitaxy (PA-MBE) under Ga-rich conditions at a temperature range of 400 °C - 800 °C. In samples with low Bi content, there is a comparatively weak subband-gap absorption with a low energy tail.\cite{levander2010gan1} The emission near the GaN band edge might be related to isoelectronic dopant characteristics in the GaN lattice or can also be attributed to cubic GaN formation as well as donor-acceptor pair (DAP) recombination.\cite{novikov2003bismuth} Moreover, Novikov \textit{et al.} and Weber \textit{et al.}\cite{novikov2012molecular, liliental2013microstructure} demonstrated strong optical absorption at a lower onset energy of $\sim2.5$ eV for a Bi concentration of $\sim5\%$. In these studies, the Ga$\rm N_{1-x}Bi_x$ nanostructures were grown by MBE on sapphire substrates under Ga-rich conditions at low temperature (80 °C - 100 °C). The enhanced absorption at lower energies with increasing Bi concentration shows that increasing substitutional Bi concentrations lead to a higher density of states as well as a larger width of the Bi-derived valence band in the pseudo-amorphous Ga$\rm N_{1-x}Bi_x$ matrix.\cite{levander2010gan1} On the other hand, Weber \textit{et al.}\cite{liliental2012structural} reported even lower energy absorption onsets of 1.0$\sim$2.0 eV for the Ga$\rm N_{1-x}Bi_x$ samples with up to $5.50\%$ Bi incorporation, similarly grown by MBE on $\rm Al_2O_3$ under different Ga:N ratios at a low temperature ($\sim100$ °C). The optical properties of these alloys may be attributed to the presence of small crystallites in the amorphous matrix. More recently, Navid \textit{et al.}\cite{navid2024structural} grew high-quality dilute Bi-doped GaN nanostructures through PA-MBE, with comprehensive structural and optical characterization, including a micro-Raman spectra analysis that suggests Bi substitutions on N sites under N-rich growth conditions.

On the theory side, previous computational studies investigating Ga$\rm N_{1-x}Bi_x$ systems are mostly at the level of the local density approximation (LDA)\cite{yan2014first} and the generalized gradient approximation (GGA)\cite{mbarki2012first} for the exchange-correlation potential of density functional theory (DFT), which are affected by the band-gap underestimation problem. Relevant thermodynamic calculations of defective Bi incorporation in GaN by Bernadini \textit{et al.}\cite{bernardini2003bi} revealed the formation of multiple-donor $\rm Bi_N$ defect hindering efficient p-type doping, and isoelectric $\rm Bi_{Ga}$ under n-type and moderate p-type conditions. However, these calculations used the GGA approximation, which underestimates the band gap, and the study did not elaborate on the band-structure properties. This motivates a systematic study of the band structure and defect properties of Bi-incorporated GaN that overcomes the band-gap problem of local DFT functionals.

In this work, we apply hybrid DFT calculations to investigate the electronic and defect properties of dilute Bi-doped GaN and understand the atomistic origin of its characteristic PL and optical absorption energies. Our studies of Bi-related defects uncover complexes between Bi incorporated on the N site with Ga vacancies in the GaN host ($\rm Bi_N + V_{Ga}$) that are found to be stable in the 3+, 0, and 3– states under varying Fermi-level positions. Our band structure calculations for these defects and complexes agree with and explain the experimental observations regarding the origins of the PL peaks and optical absorption edges. Our results clarify the defect properties of Bi-doped GaN, and outline practical strategies for controlling the site of Bi incorporation.

We employ the projector-augmented wave (PAW) method as implemented in Vienna Ab initio Simulation Package (VASP)\cite{kresse1996efficient} to perform DFT calculations with the Heyd-Scuseria-Ernzerhof (HSE) hybrid functional.\cite{heyd2003hybrid} The supercell size for Bi-related point defects in GaN is 96-atom within orthorhombic symmetry. The mixing parameter $\alpha$ for HSE is set to 0.3 to simultaneously reproduce the lattice constants (a = 3.21 \text{\AA}, c = 5.20 \text{\AA} after structural relaxations) and band gap of $\sim$3.44 eV for pristine GaN, as we have previously applied for dilute Sb-incorporated GaN.\cite{liu2024selective} A plane-wave cutoff energy of 500 eV was used, and the Brillouin zone was sampled using a $\Gamma$-centered Monkhorst-Pack k-point mesh of 2 $\times$ 2 $\times$ 2. The atomic configurations within the supercell containing the Bi-related point defects were relaxed until the residual forces were converged below 0.01 eV/\text{\AA}. Spin polarization was explicitly considered for all defect calculations. The formation energies of Bi-related defects for various charge states were evaluated as a function of Fermi level position with proper chemical potentials using the established methodology outlined by Freysoldt \textit{et al.}\cite{freysoldt2014first} The relevant chemical potentials terms, such as $\mu_{Ga}$  and $\mu_N$, are referenced to the energy per atom of bulk Ga and $\rm N_2$ molecules, respectively, and their values vary within a range defined by the DFT-calculated formation enthalpy of GaN ($\Delta H_f$(GaN) = –2.578 eV) from N-rich to Ga-rich. Similarly, $\mu_{Bi}$ is determined to bulk Bi while limited by $\Delta H_f$(GaBi) = –0.902 eV for different growth conditions.

\begin{figure}[htbp]
    \centering
    \includegraphics[width=\columnwidth]{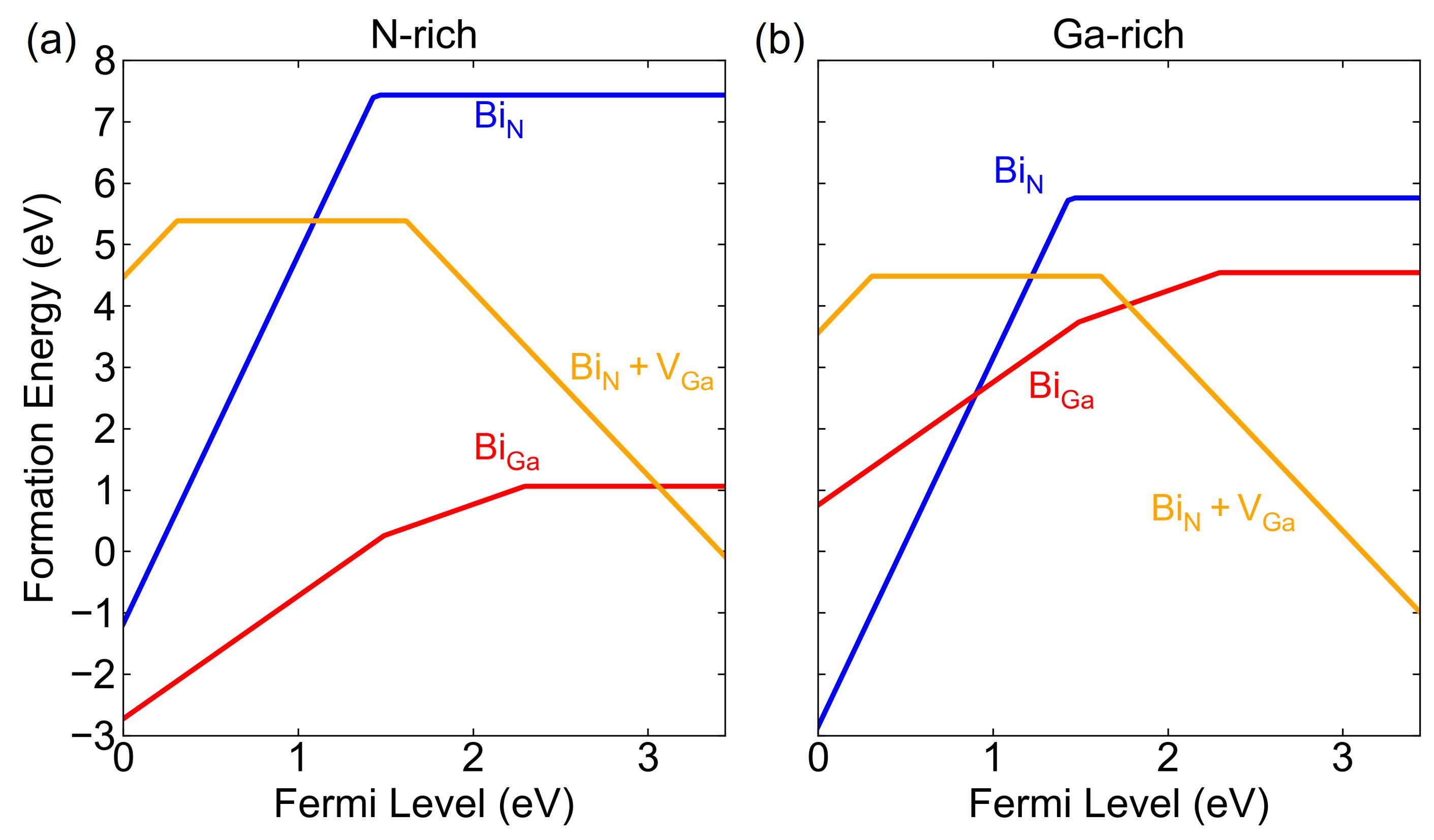}
    \caption{Formation energies for various Bi substitutions and relevant defect complexes in GaN as a function of Fermi level under (a) N-rich and (b) Ga-rich growth conditions. Both $\rm Bi_N$ and $\rm Bi_{Ga}$ act as deep donors. However, since $\rm Bi_N$ can donate up to 6 electrons and become 6+, the $\rm Bi_N + V_{Ga}$ defect complex can be thermodynamically stabilized due to charge compensation, particularly under n-type conditions.}
    \label{fig:1}
\end{figure}

We first examine the formation energies of Bi-related substitutional defects and defect complexes in GaN as shown in Fig. \ref{fig:1}. Similar to the results previously reported by Bernardini \textit{et al.}\cite{bernardini2003bi}, cationic $\rm Bi_{Ga}$ adopts the charge states of +2, +1, and 0, while anionic $\rm Bi_N$ occurs as +6 and 0 for Fermi level values near the band edges and as +1 in a very short range near the middle of the gap. The charge-transition levels are listed in Table I. We find that $\rm Bi_{Ga}$ dominates under N-rich conditions. However, the difference of their neutral-state formation energies becomes much smaller ($\sim$1 eV), making the two substitutions more comparable in energy under Ga-rich conditions. Moreover, $\rm Bi_{Ga}$ is energetically favored over $\rm Bi_N$ for $\rm E_F$ above 0.90 eV, and acts as a deep double donor. For $\rm E_F$ between 1.49 eV and 2.30 eV, $\rm Bi_{Ga}$ behaves as a single donor, while a competition between $\rm Bi_{Ga}^0$ and $\rm Bi_N^0$ emerges under p-type conditions. On the other hand, unlike other group-V elements such as As and Sb, $\rm Bi_N$ can donate up to 6 electrons, three of which come from the 6p orbital of Bi, which becomes empty, and three from the N vacancy needed to accommodate the N-substitutional Bi ion. Therefore, the incorporation of large lattice-mismatched $\rm Bi_N$ seems to be limited to p-type conditions, while the excess electrons from $\rm Bi_N^{6+}$ shift the Fermi level towards the conduction band and induce strong self-compensation effects.

\begin{table}[htbp]
    \centering
    \begin{tabular}{c|c|c}
         \textbf{Defect} &  \textbf{Transition} & \textbf{Energy (eV)}  \\
         \hline
         $\rm Bi_{Ga}$  & (+2/+1)  & 1.49 \\
          $\rm Bi_{Ga}$  & (+1/0)  & 2.30 \\
          $\rm Bi_{N}$  & (+6/+1)  & 1.43 \\
          $\rm Bi_{N}$  & (+1/0)  &  1.47 \\
          $\rm Bi_{N} + V_{Ga}$  & (0/-3)  & 0.31 \\
          $\rm Bi_N + V_{Ga}$  & (+3/0)  & 1.62 \\
    \end{tabular}
    \caption{Table I. Thermodynamic charge-transition levels of $\rm Bi_N$, $\rm Bi_{Ga}$, and $\rm Bi_N + V_{Ga}$.}
    \label{tab:table}
\end{table}

However, Bi can also be incorporated in the N site due to the formation of a defect complex with Ga vacancies, $\rm Bi_N + V_{Ga}$. This complex is stable in the –3 charge state for Fermi levels near the conduction band minimum (CBM) and shows lower formation energy compared to substitutional $\rm Bi_{Ga}$ even under N-rich conditions. The stability of this $\rm Bi_N + V_{Ga}$ complex can be explained based on the fact that $\rm Bi_N$ can donate up to 6 electrons, making it electronically attracted to negatively charged $\rm V_{Ga}$. The $\rm Bi_N + V_{Ga}$ DAP stability is unique to Bi in GaN, since this complex is unstable and cannot form for other group-V donors such as As and Sb. This special type of defect complex leads us to reexamine the incorporating of Bi in GaN, and sheds light on the possible origin of the unexplained PL and absorption peaks in experiments.

\begin{figure*}[htbp]
    \centering
    \includegraphics[width=1.5\columnwidth]{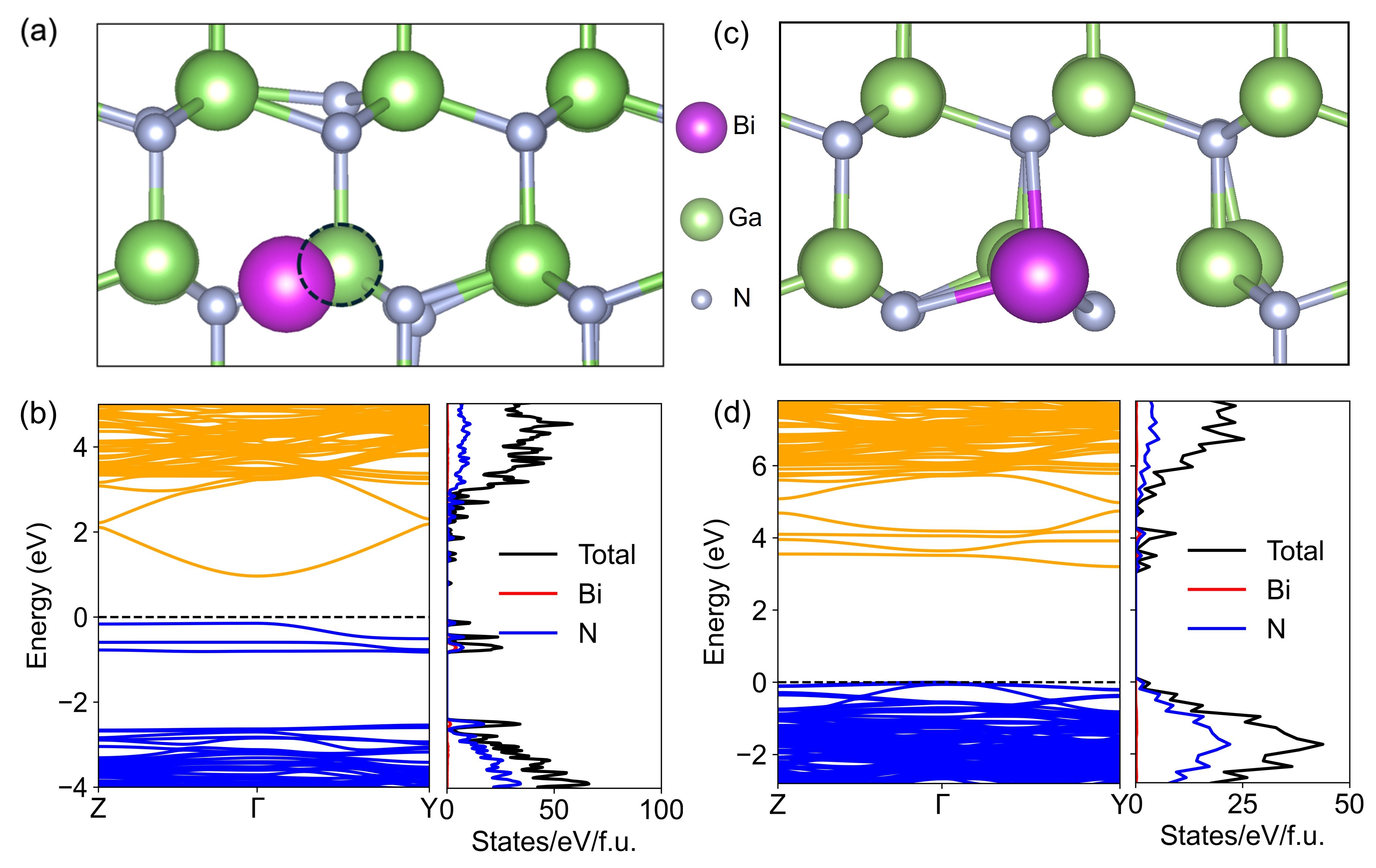}
    \caption{(a) Atomic configurations of $\rm (Bi_N + V_{Ga})^{3-}$ in GaN. The substitutional Bi atom moves towards the nearby Ga vacancy to reduce the local strain. (b) Band structure and density of states (DOSs) of $\rm (Bi_N + V_{Ga})^{3-}$ system, hosting a band gap $\sim$1.11 eV, i.e., close to the lowest absorption tail reported by Weber \textit{et al.}\cite{liliental2012structural} (c) Atomic configurations of $\rm (Bi_N + V_{Ga})^{3+}$ in GaN. Bi atom now moves further towards the Ga vacancy site. (d) Band structure and DOSs of $\rm (Bi_N + V_{Ga})^{3+}$ system, hosting a band gap $\sim$3.17 eV, aligned with the near GaN band edge absorption reported by Levander \textit{et al.}\cite{levander2010gan1} and Novikov \textit{et al.}\cite{novikov2012molecular}}
    \label{fig:2}
\end{figure*}

To elucidate the structural and electronic properties of $\rm Bi_N + V_{Ga}$, we perform structural relaxations for this point defect complex in GaN to analyze the atomic configurations and the corresponding band structures in its +3 and –3 charge states in Fig. \ref{fig:2}. As shown in the structural model in Fig. \ref{fig:2} (a) for $(\rm Bi_N + V_{Ga})^{3-}$, the Bi atom lies in the space in-between the substitutional N and the vacant Ga sites, due to its relatively large atomic radius that is highly lattice mismatched to the interatomic spacing in GaN. The band structures and the atomic-projected density of states (pDOS) in Fig. \ref{fig:2} (b) elucidate that $(\rm Bi_N + V_{Ga})^{3-}$ gives rise to occupied mid-gap states arising from the Bi 6p orbital, severely reducing the band gap to $\sim$1.11 eV. This band-gap value agrees with the lowest energy absorption tail observed by Weber \textit{et al.}\cite{liliental2012structural} Introducing the vacancy of Ga as a DAP with Bi also serves to release the local strain induced by isolated Bi incorporation in the substitutional sites. Furthermore, the atomic geometry of $(\rm Bi_N + V_{Ga})^{3+}$ in Fig. \ref{fig:2} (c) shows that Bi occupies the Ga vacancy position instead of directly substituting N, and Fig. \ref{fig:2} (d) exhibits near-CBM shallow donor contributions from Bi-related defect bands as Bi 6p orbital becomes empty. The total band gap of $\sim$3.17 eV agrees with the PL peak positions and the absorption onset at $\sim$3.2 eV reported by Levander \textit{et al.}\cite{levander2010gan1} and Novikov \textit{et al.}\cite{novikov2012molecular}

\begin{figure*}
    \centering
    \includegraphics[width=2.0\columnwidth]{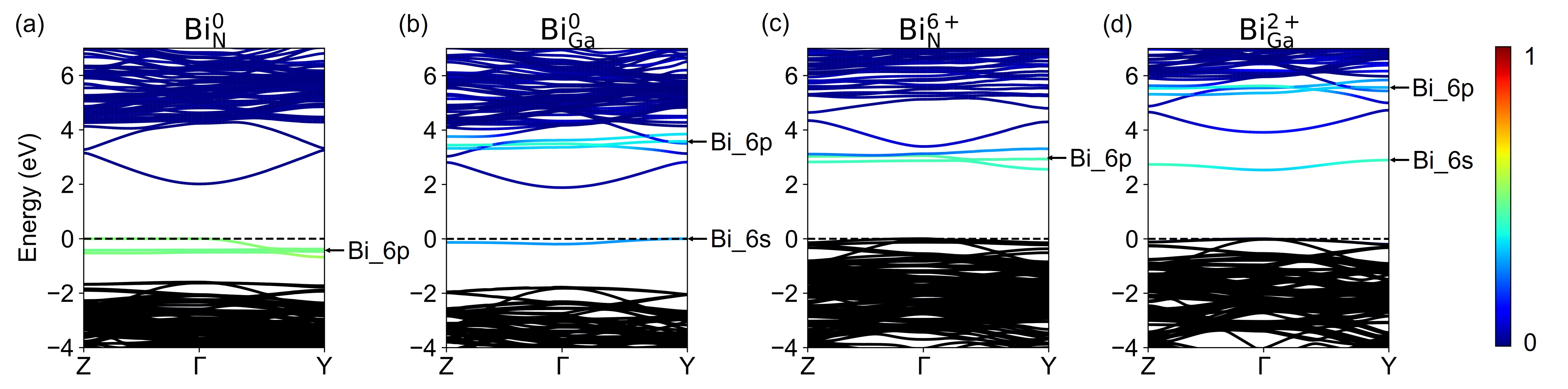}
    \caption{Band structures of different Bi substitutions into GaN. (a) $\rm Bi_N^0$ with a band gap of 2.01 eV. (b) $\rm Bi_{Ga}^0$ with a band gap of 1.88 eV. These two values are aligned with the absorption reported by Weber \textit{et al.}\cite{liliental2012structural} (c) $\rm Bi_N^{6+}$ with a band gap of 2.55 eV, and (d) $\rm Bi_{Ga}^{2+}$ with a band gap of 2.53 eV, consistent with the absorption results reported by Novikov \textit{et al.}\cite{zhang2002influence} and Weber \textit{et al.}\cite{ibanez2005optical, kudrawiec2020bandgap}}
    \label{fig:3}
\end{figure*}

In addition to the above theoretical clarifications about previously reported experimental near-ultraviolet and near-infrared region for Bi-doped GaN, we further investigate the band structures for all possible cationic and anionic substitutional Bi defects in their available charge states to provide a complete and consistent explanation of the other reported band-gap values of $\sim$2.0 to 2.5 eV in the visible range. Our calculated results are shown in Fig. \ref{fig:3}. For charge-neutral substitutions, i.e., $\rm Bi_N^0$ and $\rm Bi_{Ga}^0$ in Fig. \ref{fig:3} (a) and (b), although we obtain similar band gaps of 2.01 eV and 1.88 eV, we find that these values have completely different chemical origins. For example, $\rm Bi_N^0$ behaves as an anion with fully occupied 6s and 6p orbitals, and the relevant bands lie within the gap and below the Fermi level. In contrast, the trivalent Bi atom in $\rm Bi_{Ga}^0$ exhibits a fully filled 6s but an empty 6p orbital, giving rise to three p states above the CBM and one occupied midgap s state. Experimentally, these two Bi neutral cationic and anionic substitutions are indistinguishable solely from emission and absorption spectra. Moreover, they can coexist under Ga-rich growth conditions since their formation energies differ by only $\sim$1 eV. We have previously reported similarly the coexistence of cationic and anionic Sb substitutions in Sb-doped GaN.\cite{liu2024selective} Moreover, given that $\rm Bi_N$ and $\rm Bi_{Ga}$ can also exist in their positively charged states ($\rm Bi_N^{6+}$ and $\rm Bi_{Ga}^{2+}$), we plot the corresponding band structures in Figs. \ref{fig:3} (c) and (d). In these two cases, the orbitals of the sixth shell of Bi occur near the CBM, and exhibit similar band gap values of 2.55 eV and 2.53 eV, which are in good agreement with the measurements of Novikov \textit{et al.}\cite{novikov2012molecular} and Weber \textit{et al.}\cite{liliental2013microstructure} 

Our findings that Bi incorporation in GaN significantly influences its structural and electronic properties underscore the importance of controlling cation incorporation in highly mismatched nitride alloys in order to enable device applications. Bi can adopt different incorporation sites and charge states, resulting in a range of band gaps, which offers tunability for various optoelectronic devices. Meanwhile, the controllability of Bi incorporation sites presents an opportunity to engineer material properties by controlling growth conditions, such as temperature, pressure, and precursor flow rate to tune the chemical potential during epitaxial deposition. Additionally, the doping type (n-type vs. p-type) and growth conditions (N-rich v.s. Ga-rich) play a critical role in stabilizing the various Bi configurations, which can be leveraged to tailor the band gap and carrier dynamics. Under Ga-rich n-type conditions, $\rm Bi_N^0$ and $\rm Bi_{Ga}^0$ are indistinguishable solely from the absorption peak because of their similar formation energies and band gap values, but p-type doping favors the stability of $\rm Bi_N^{6+}$. For N-rich conditions, n-type doping gives rise to $\rm (Bi_N + V_{Ga})^{3-}$, while the neutral and positively-charged $\rm Bi_{Ga}$ become stable at lower $\rm E_F$. This site-specific control of Bi incorporation enables the design of GaN-based materials with precisely tuned band gaps for applications in NIR photodetectors, LEDs, and high-efficiency photovoltaics. 

In summary, we apply first-principles calculations to investigate the origin of the various experimental PL peaks and optical absorption onsets in dilute Bi-doped GaN. The thermodynamic properties of substitutional Bi indicate the stability as well as the tunability of dilute Bi incorporation on the Ga site as a cation and on the N site as an anion. Moreover, we identify the charge-compensated defect complex $\rm Bi_N + V_{Ga}$ as a stable compensating defect both under n-type and under p-type conditions. Our band-structure calculations further confirm the role of various Bi incorporations in explaining available observations in optical experiments. Our work highlights the importance of understanding and controlling the incorporation of highly mismatched elements into GaN in order to expand its functionality for next-generation semiconductor devices. 

\begin{acknowledgments}
This work was supported by the U.S. Army Research Office Award No. W911NF2110337. Computational resources were provided by the National Energy Research Scientific Computing Center, which is supported by the Office of Science of the U.S. Department of Energy under Contract No. DE-AC02-05CH11231.
\end{acknowledgments}

\nocite{*}
\bibliography{aipsamp}

\end{document}